\documentclass[12pt]{article}
\usepackage{lineno}  
\usepackage{mciteplus}
\usepackage{graphicx}
\usepackage{ifthen} 
\usepackage{amsmath} 
\usepackage{amssymb}
\usepackage{amsfonts}
\usepackage{hyperref}    
\usepackage{cite} 
\usepackage{xspace}
\usepackage{pstricks}
\newboolean{uprightparticles}
\setboolean{uprightparticles}{false} 
\newboolean{inbibliography}
\setboolean{inbibliography}{false} 
\newboolean{articletitles}
\setboolean{articletitles}{true} 

\def\dzero  {\mbox{D0}\xspace}

\def\MagUp {\mbox{\em Mag\kern -0.05em Up}\xspace}

\ifthenelse{\boolean{uprightparticles}}%
{

 \def\Pmu         {\ensuremath{\upmu}\xspace}

 \def\Ppi         {\ensuremath{\uppi}\xspace}

 \def\Ppsi        {\ensuremath{\uppsi}\xspace}

 \def\PDelta      {\ensuremath{\Delta}\xspace}                 
 \def\PXi      {\ensuremath{\Xi}\xspace}                 
 \def\PLambda      {\ensuremath{\Lambda}\xspace}                 
 \def\PSigma      {\ensuremath{\Sigma}\xspace}                 
 \def\POmega      {\ensuremath{\Omega}\xspace}                 
 \def\PUpsilon      {\ensuremath{\Upsilon}\xspace}                 
 
 \def\PB      {\ensuremath{\mathrm{B}}\xspace}                 
                  
 \def\PD      {\ensuremath{\mathrm{D}}\xspace}

 \def\PJ      {\ensuremath{\mathrm{J}}\xspace}                 
 \def\PK      {\ensuremath{\mathrm{K}}\xspace}

 \def\PW      {\ensuremath{\mathrm{W}}\xspace}

 \def\Pb      {\ensuremath{\mathrm{b}}\xspace}                 
 \def\Pc      {\ensuremath{\mathrm{c}}\xspace}                 
                  
 \def\Pe      {\ensuremath{\mathrm{e}}\xspace}

 \def\Pi      {\ensuremath{\mathrm{i}}\xspace}

 \def\Pp      {\ensuremath{\mathrm{p}}\xspace}

 \def\Ps      {\ensuremath{\mathrm{s}}\xspace}                 
 \def\Pt      {\ensuremath{\mathrm{t}}\xspace}                 
 \def\Pu      {\ensuremath{\mathrm{u}}\xspace}

}
{

 \def\Pmu         {\ensuremath{\mu}\xspace}

 \def\Ppi         {\ensuremath{\pi}\xspace}

 \def\Ppsi        {\ensuremath{\psi}\xspace}                 
                  
 \mathchardef\PDelta="7101
 \mathchardef\PXi="7104
 \mathchardef\PLambda="7103
 \mathchardef\PSigma="7106
 \mathchardef\POmega="710A
 \mathchardef\PUpsilon="7107
                  
 \def\PB      {\ensuremath{B}\xspace}                 
                  
 \def\PD      {\ensuremath{D}\xspace}

 \def\PJ      {\ensuremath{J}\xspace}                 
 \def\PK      {\ensuremath{K}\xspace}

 \def\PW      {\ensuremath{W}\xspace}

 \def\Pb      {\ensuremath{b}\xspace}                 
 \def\Pc      {\ensuremath{c}\xspace}                 
                  
 \def\Pe      {\ensuremath{e}\xspace}

 \def\Pi      {\ensuremath{i}\xspace}

 \def\Pp      {\ensuremath{p}\xspace}

 \def\Ps      {\ensuremath{s}\xspace}                 
 \def\Pt      {\ensuremath{t}\xspace}                 
 \def\Pu      {\ensuremath{u}\xspace}

}

\makeatletter
\ifcase \@ptsize \relax
  \newcommand{\miniscule}{\@setfontsize\miniscule{4}{5}}
\or
  \newcommand{\miniscule}{\@setfontsize\miniscule{5}{6}}
\or
  \newcommand{\miniscule}{\@setfontsize\miniscule{5}{6}}
\fi
\makeatother

\DeclareRobustCommand{\optbar}[1]{\shortstack{{\miniscule (\rule[.5ex]{1.25em}{.18mm})}
  \\ [-.7ex] $#1$}}

\def\epem       {{\ensuremath{\Pe^+\Pe^-}}\xspace}

\def\mup        {{\ensuremath{\Pmu^+}}\xspace}
\def\mun        {{\ensuremath{\Pmu^-}}\xspace} 
\def\mumu       {{\ensuremath{\Pmu^+\Pmu^-}}\xspace}

\def\ellm       {{\ensuremath{\ell^-}}\xspace}
\def\ellp       {{\ensuremath{\ell^+}}\xspace}

\def\W      {{\ensuremath{\PW}}\xspace}

\def\uquark    {{\ensuremath{\Pu}}\xspace}

\def\squark    {{\ensuremath{\Ps}}\xspace}
\def\squarkbar {{\ensuremath{\overline \squark}}\xspace}

\def\cquark    {{\ensuremath{\Pc}}\xspace}
\def\cquarkbar {{\ensuremath{\overline \cquark}}\xspace}

\def\bquark    {{\ensuremath{\Pb}}\xspace}
\def\bquarkbar {{\ensuremath{\overline \bquark}}\xspace}

\def\tquark    {{\ensuremath{\Pt}}\xspace}

\def\pion   {{\ensuremath{\Ppi}}\xspace}

\def\pip    {{\ensuremath{\pion^+}}\xspace}
\def\pim    {{\ensuremath{\pion^-}}\xspace}
\def\pipm   {{\ensuremath{\pion^\pm}}\xspace}

\def\kaon    {{\ensuremath{\PK}}\xspace}
  \def\Kbar    {{\kern 0.2em\overline{\kern -0.2em \PK}{}}\xspace}

\def\KorKbar    {\kern 0.18em\optbar{\kern -0.18em K}{}\xspace}
\def\Kz      {{\ensuremath{\kaon^0}}\xspace}

\def\Kp      {{\ensuremath{\kaon^+}}\xspace}
\def\Km      {{\ensuremath{\kaon^-}}\xspace}
\def\Kpm     {{\ensuremath{\kaon^\pm}}\xspace}

\def\KS      {{\ensuremath{\kaon^0_{\rm\scriptscriptstyle S}}}\xspace}

\def\Kstarz  {{\ensuremath{\kaon^{*0}}}\xspace}
\def\Kstarzb {{\ensuremath{\Kbar{}^{*0}}}\xspace}
\def\Kstar   {{\ensuremath{\kaon^*}}\xspace}

\def\Dbar    {{\kern 0.2em\overline{\kern -0.2em \PD}{}}\xspace}
\def\D       {{\ensuremath{\PD}}\xspace}

\def\DorDbar    {\kern 0.18em\optbar{\kern -0.18em D}{}\xspace}
\def\Dz      {{\ensuremath{\D^0}}\xspace}
\def\Dzb     {{\ensuremath{\Dbar{}^0}}\xspace}

\def\Ds      {{\ensuremath{\D^+_\squark}}\xspace}
\def\Dsp     {{\ensuremath{\D^+_\squark}}\xspace}
\def\Dsm     {{\ensuremath{\D^-_\squark}}\xspace}
\def\Dspm    {{\ensuremath{\D^{\pm}_\squark}}\xspace}

\def\B       {{\ensuremath{\PB}}\xspace}
\def\Bbar    {{\ensuremath{\kern 0.18em\overline{\kern -0.18em \PB}{}}}\xspace}

\def\BorBbar    {\kern 0.18em\optbar{\kern -0.18em B}{}\xspace}
\def\Bz      {{\ensuremath{\B^0}}\xspace}

\def\Bu      {{\ensuremath{\B^+}}\xspace}
\def\Bub     {{\ensuremath{\B^-}}\xspace}
\def\Bp      {{\ensuremath{\Bu}}\xspace}
\def\Bm      {{\ensuremath{\Bub}}\xspace}
\def\Bpm     {{\ensuremath{\B^\pm}}\xspace}

\def\Bd      {{\ensuremath{\B^0}}\xspace}
\def\Bs      {{\ensuremath{\B^0_\squark}}\xspace}

\def\jpsi     {{\ensuremath{{\PJ\mskip -3mu/\mskip -2mu\Ppsi\mskip 2mu}}}\xspace}

  \def\Y#1S{\ensuremath{\PUpsilon{(#1S)}}\xspace}

\def\proton      {{\ensuremath{\Pp}}\xspace}
\def\antiproton  {{\ensuremath{\overline \proton}}\xspace}

\def\Lz          {{\ensuremath{\PLambda}}\xspace}
\def\Lbar        {{\ensuremath{\kern 0.1em\overline{\kern -0.1em\PLambda}}}\xspace}
\def\LorLbar    {\kern 0.18em\optbar{\kern -0.18em \PLambda}{}\xspace}

\def\Lb      {{\ensuremath{\Lz^0_\bquark}}\xspace}

\newcommand{\decay}[2]{\ensuremath{#1\!\to #2}\xspace}         

\def\to                 {\ensuremath{\rightarrow}\xspace}

\def\CP                {{\ensuremath{C\!P}}\xspace}

\def\Vcs  {{\ensuremath{V_{\cquark\squark}}}\xspace}
\def\Vts  {{\ensuremath{V_{\tquark\squark}}}\xspace}

\def\Vcbs  {{\ensuremath{V_{\cquark\bquark}^\ast}}\xspace}
\def\Vtbs  {{\ensuremath{V_{\tquark\bquark}^\ast}}\xspace}

\def\AT#1     {\ensuremath{A_{\mathrm{T}}^{#1}}\xspace}           

\def\Bsmm     {\decay{\Bs}{\mup\mun}}
\def\Bdmm     {\decay{\Bd}{\mup\mun}}

\def\C#1      {\ensuremath{\mathcal{C}_{#1}}\xspace}                       
\def\Cp#1     {\ensuremath{\mathcal{C}_{#1}^{'}}\xspace}                    
\def\Ceff#1   {\ensuremath{\mathcal{C}_{#1}^{\mathrm{(eff)}}}\xspace}        
\def\Cpeff#1  {\ensuremath{\mathcal{C}_{#1}^{'\mathrm{(eff)}}}\xspace}       
\def\Ope#1    {\ensuremath{\mathcal{O}_{#1}}\xspace}                       
\def\Opep#1   {\ensuremath{\mathcal{O}_{#1}^{'}}\xspace}                    

\newcommand{\tev}{\ifthenelse{\boolean{inbibliography}}{\ensuremath{~T\kern -0.05em eV}\xspace}{\ensuremath{\mathrm{\,Te\kern -0.1em V}}}\xspace}
\newcommand{\gev}{\ensuremath{\mathrm{\,Ge\kern -0.1em V}}\xspace}
\newcommand{\mev}{\ensuremath{\mathrm{\,Me\kern -0.1em V}}\xspace}
\newcommand{\kev}{\ensuremath{\mathrm{\,ke\kern -0.1em V}}\xspace}
\newcommand{\ev}{\ensuremath{\mathrm{\,e\kern -0.1em V}}\xspace}
\newcommand{\gevc}{\ensuremath{{\mathrm{\,Ge\kern -0.1em V\!/}c}}\xspace}
\newcommand{\mevc}{\ensuremath{{\mathrm{\,Me\kern -0.1em V\!/}c}}\xspace}
\newcommand{\gevcc}{\ensuremath{{\mathrm{\,Ge\kern -0.1em V\!/}c^2}}\xspace}
\newcommand{\gevgevcccc}{\ensuremath{{\mathrm{\,Ge\kern -0.1em V^2\!/}c^4}}\xspace}
\newcommand{\mevcc}{\ensuremath{{\mathrm{\,Me\kern -0.1em V\!/}c^2}}\xspace}

\def\invfb   {\ensuremath{\mbox{\,fb}^{-1}}\xspace}

\def\gsim{{~\raise.15em\hbox{$>$}\kern-.85em
          \lower.35em\hbox{$\sim$}~}\xspace}
\def\lsim{{~\raise.15em\hbox{$<$}\kern-.85em
          \lower.35em\hbox{$\sim$}~}\xspace}

\def\degrees{\ensuremath{^{\circ}}\xspace}

\def\rad{\ensuremath{\rm \,rad}\xspace}

\def\tell1  {TELL1\xspace}
\def\ukl1   {UKL1\xspace}

\def\ellell{\ensuremath{\ell^+\ell^-}\xspace}
\usepackage{dcolumn}
\newcolumntype{C}{>{$}c<{$}}
\newcolumntype{R}{>{$}r<{$}}
\newcolumntype{L}{>{$}l<{$}}
\newcommand{\aerr}[2]{{\:}^{+{\:}#1}_{-{\:}#2}}%

\newcommand\pubnumber{SNSN-323-63}
\newcommand\pubdate{\today}

\def\Title#1{\begin{center} {\Large #1 } \end{center}}
\def\Author#1{\begin{center}{ \sc #1} \end{center}}
\def\Address#1{\begin{center}{ \it #1} \end{center}}

\newcommand\pubblock{\rightline{\begin{tabular}{l} \pubnumber\\
         \pubdate  \end{tabular}}}
\newenvironment{Abstract}{\begin{quotation}  }{\end{quotation}}
\newenvironment{Presented}{\begin{quotation} \begin{center} 
             PRESENTED AT\end{center}\bigskip 
      \begin{center}\begin{large}}{\end{large}\end{center} \end{quotation}}

\begin{document}
\begin{titlepage}
\pubblock

\vfill
\Title{CKM studies from \bquark physics at hadron machines}
\vfill
\Author{Patrick Koppenburg}
\Address{Nikhef, Science Park 105\\
1098 XG, Amsterdam, The Netherlands}
\vfill
\begin{Abstract}
In absence of direct signs of new physics at the LHC,
flavour physics provides an ideal laboratory to 
look for deviations from the Standard Model and explore an energy regime
beyond the LHC reach. Here, new results in \CP violation
and rare decays are presented.
\end{Abstract}
\vfill
\begin{Presented}
CKM workshop 2014, Vienna
\end{Presented}
\vfill
\end{titlepage}
\def\thefootnote{\fnsymbol{footnote}}
\setcounter{footnote}{0}
%

\section{Introduction}
The first run of the Large Hadron Collider with 7 and 8\:\tev $pp$ collisions has allowed
to discover the Higgs boson~\cite{Aad:2012tfa,*Chatrchyan:2012ufa}, but not to find any hint of the existence of
new particles. Neither supersymmetry nor any other sign of new
physics has popped out of the LHC. This situation may change during Run II
with higher centre-of-mass energies of 13\:\tev. In the meantime
it is worth investigating what the data at the lower energies involved in
the decays of \bquark hadrons can tell us about new physics.

Studies of \CP violation in heavy flavour decays are both sensitive 
to the above-mentioned
high mass scales and to potential new phases beyond the phase of the CKM matrix. 
Also of particular interest are rare decays that are strongly suppressed in the Standard Model (SM),
where new physics amplitudes could be sizable. 
\section{\CP violation measurements}\label{Sec:CPV}
Owing to the legacy of the \B factories~\cite{Bevan:2014iga}, we have entered 
the era of precision tests of the Cabibbo-Kobayashi-Maskawa (CKM) paradigm~\cite{Kobayashi:1973fv}.
Precise measurements of the angles of the unitarity triangle(s) are needed 
to search for new sources of \CP violation beyond the single phase of the CKM matrix.

\begin{figure}[tb]
\includegraphics[height=0.39\textwidth]{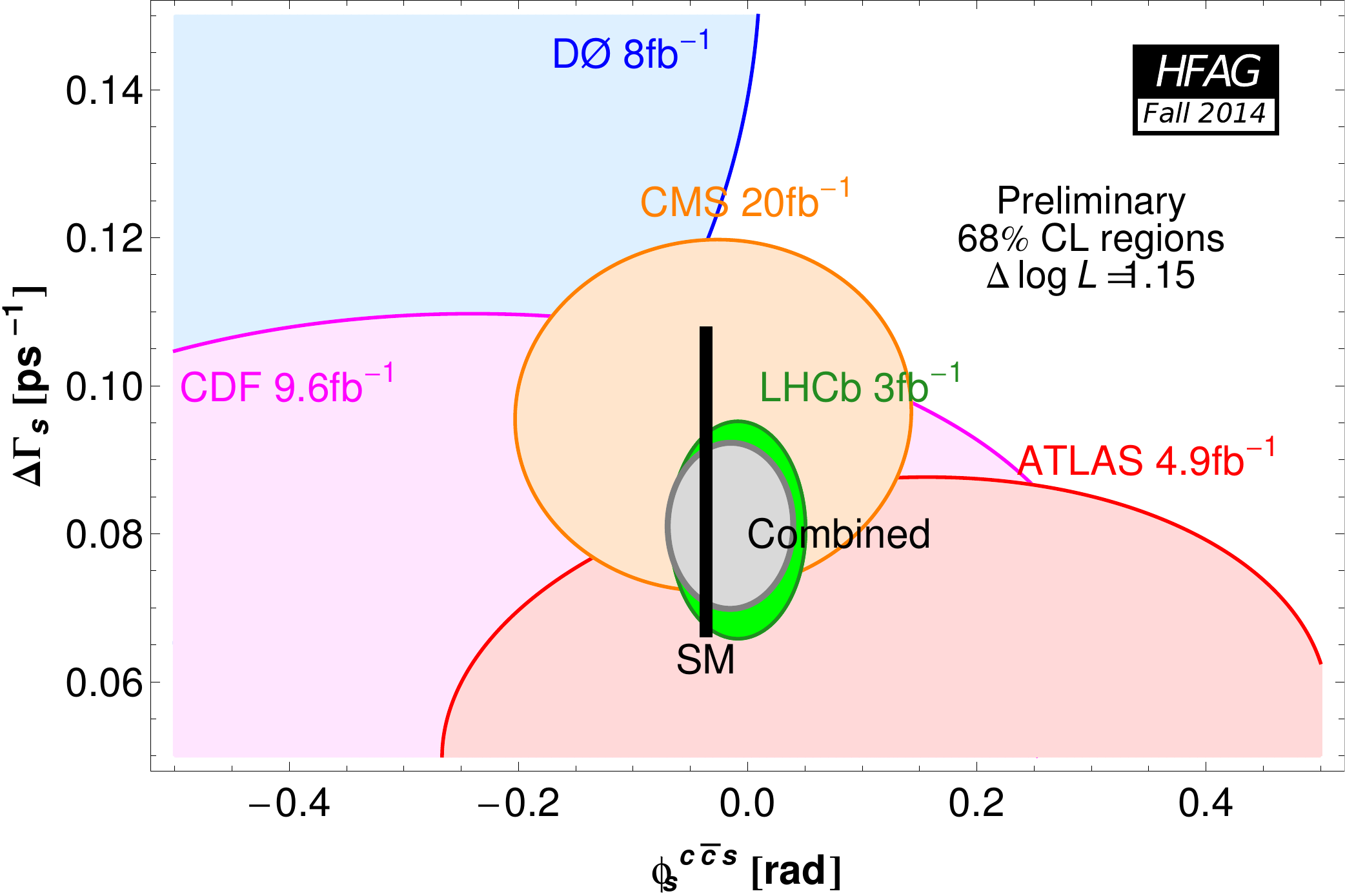}\hskip 0.02\textwidth 
\includegraphics[height=0.39\textwidth]{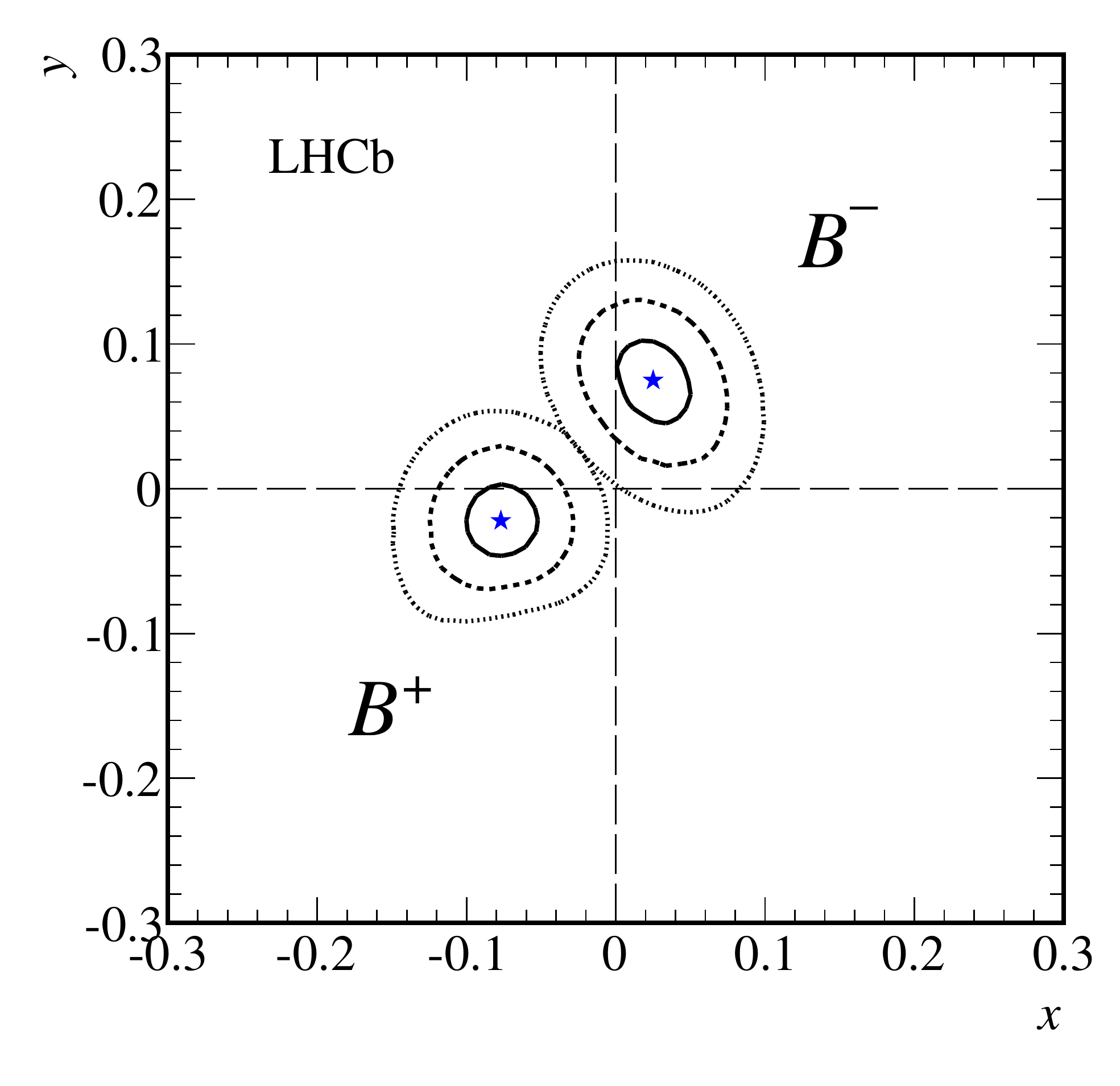}
\caption{(left) Constraints on $\Delta\Gamma_s$ and 
  $\varphi_s^{\cquark\cquarkbar\squark}$ from various
  experiments. (right) Fitted values of $x$ and $y$ for \Bm and 
  \Bp in \decay{\Bpm}{\D\Kpm} with \decay{\D}{\KS h^+h^-} decays~\cite{LHCb-PAPER-2014-041}.}
\label{Fig:2014-059}\label{Fig:2014-041}
\end{figure}

The LHC is often considered as a \Bs meson factory, owing to its large
cross-section and the unprecedented capabilities of the LHC experiments to precisely
resolve its oscillations. This opens the door to precision measurements of the \CP-violating
phase $\varphi_s^{\cquark\cquarkbar\squark}$, which is equal to 
$-2\beta_s\equiv-2\arg\left(-\Vts\Vtbs/\Vcs\Vcbs\right)=-0.0363\pm0.0013$ in the SM,
neglecting sub-leading penguin contributions.
It was measured at the LHC 
using the flavour eigenstate decays \decay{\Bs}{\jpsi\phi} with \decay{\jpsi}{\mumu} and 
\decay{\phi}{\Kp\Km}~\cite{Aad:2014cqa,*CMS-PAS-BPH-13-012,*LHCb-PAPER-2013-002} and 
\decay{\Bs}{\jpsi\pip\pim}~\cite{LHCb-PAPER-2014-019}. Recently LHCb used the decay 
\decay{\Bs}{\jpsi\Kp\Km} for the first time in a polarisation-dependent way~\cite{LHCB-PAPER-2014-059}. 
Combined with \decay{\Bs}{\jpsi\pip\pim}, LHCb obtains 
$\varphi_s^{\cquark\cquarkbar\squark}=-0.010\pm 0.039$ rad. The 
constraints on $\varphi_s^{\cquark\cquarkbar\squark}$ and the decay width difference 
$\Delta\Gamma_s$  are shown in Fig.~\ref{Fig:2014-059} (left).
The same quantity was also measured with a fully hadronic final state using the decay
\decay{\Bs}{\Dsp\Dsm} with \decay{\Dspm}{\Kp\Km\pipm}, yielding $0.02\pm0.17\pm0.02$~\cite{LHCb-PAPER-2014-051}. 
The effective tagging power $\epsilon{\cal D}^2$ in excess of 
5\% for this channel is unprecedented at a hadron collider.

With the precision on some CKM phases reaching the degree level, the effects of suppressed penguin 
topologies cannot be neglected any 
more~\cite{Fleischer:1999nz,*Fleischer:1999zi,*Fleischer:1999sj,*Faller:2008zc,*DeBruyn:2010hh,Ciuchini:2005mg}.
Cabibbo-suppressed decay modes, where these topologies are relatively more prominent
can be used to constrain such effects. This programme has started with studies of the decays 
\decay{\Bs}{\jpsi\KS}~\cite{LHCb-PAPER-2013-015},
\decay{\Bs}{\jpsi\Kstarzb}~\cite{LHCb-PAPER-2012-014} and more recently with 
\decay{\Bz}{\jpsi\pip\pim}~\cite{LHCb-PAPER-2014-058}. The measurement of 
$\sin2\beta^\text{eff}$ in the latter mode allows to constrain the 
shift to $\varphi_s^{\cquark\cquarkbar\squark}$ due to penguin topologies
to the range $[-0.018,0.021]$ radians at 68\% CL. Considering the present 
uncertainty of $\pm0.039$ radians, such a shift needs to be to constrained further.

An interesting test of the Standard Model is provided by the measurement of the mixing phase
$\varphi_s^{\squark\squarkbar\squark}$
with a purely penguin-induced mode as \decay{\Bs}{\phi\phi}.
In this case the measured value is $-0.17\pm 0.15\pm 0.03$~\cite{LHCb-PAPER-2014-026},
which is compatible with the SM expectation.

Similarly, the decays \decay{\B}{hh} with $h=\pion,\kaon$ are also sensitive to
penguin topologies (as well as trees) and are sensitive to the CKM phases 
$\gamma$ and $\beta_s$.
LHCb for the first time measured time-dependent \CP-violating observables in
\Bs decays using the decay 
\decay{\Bs}{\Kp\Km}~\cite{LHCb-PAPER-2013-040}. Using methods outlined in 
Refs.~\cite{Fleischer:1999pa,*Fleischer:2007hj,Ciuchini:2005mg}, a combination of this
and other results from \decay{\B}{hh} modes allows to determine 
$-2\beta_s = -0.12 \aerr{0.14}{0.16}$\:\rad
using as input the angle $\gamma$ from tree decays (see below), or 
$\gamma= (63.5\aerr{7.2}{6.7})^\circ$ constraining $-2\beta_s$ to the 
SM value~\cite{LHCb-PAPER-2013-045}.
These values are in principle sensitive to
the amount of U-spin breaking that is allowed in this decay and are given here 
for a maximum allowed breaking of 50\%.

This value of $\gamma$ can be compared to that obtained from tree-dominated
\decay{\B}{\D\kaon} decays, where the \CP-violating phase appears in the interference of
the \decay{\bquark}{\cquark} and \decay{\bquark}{\uquark} topologies. It is
the least precisely known angle of the unitarity matrix, and its determination
from tree decays is considered free from contributions beyond the Standard Model
and unaffected by hadronic uncertainties. Yet its precise determination is important
to test the consistency of the CKM paradigm, and to allow comparisons with determinations
from modes dominated by penguin topologies.

The most precise determination of $\gamma$ from a single decay mode is achieved with 
\decay{\Bp}{D\Kp} followed by \decay{D}{\KS h^+h^-} with 
$h=\pion,\kaon$~\cite{LHCb-PAPER-2014-041}. 
Here the interference of the \Dz and \Dzb decay to $\KS h^+h^-$ 
is exploited to measure \CP asymmetries~\cite{Giri:2003ty}. The method needs external
input in the form of a measurement of the strong phase over the Dalitz plane
of the \D decay, coming from CLEO-c data~\cite{Libby:2010nu}. The 
determined \CP-violating parameters are shown in Fig.~\ref{Fig:2014-041} (right),
and the value of $\gamma$ is $(62\aerr{15}{14})\degrees$. The same decay mode
is also used in a model-dependent measurement~\cite{LHCb-PAPER-2014-017} using
an amplitude model.

An experimentally very different way of determining $\gamma$ is provided by the decay 
\decay{\Bs}{\Ds^\pm\kaon^\mp}~\cite{Dunietz:1987bv,*Aleksan:1991nh,*Fleischer:2003yb,LHCb-PAPER-2014-038}. In this case the phase 
is measured in a time-dependent tagged \CP-violation analysis. Using a dataset 
corresponding to 1\:\invfb, LHCb determines $\gamma=(115\aerr{28}{43})\degrees$,
which is not competitive with other methods but will provide important cross-checks
 with more data.

\begin{figure}[tb]
\includegraphics[width=0.49\textwidth]{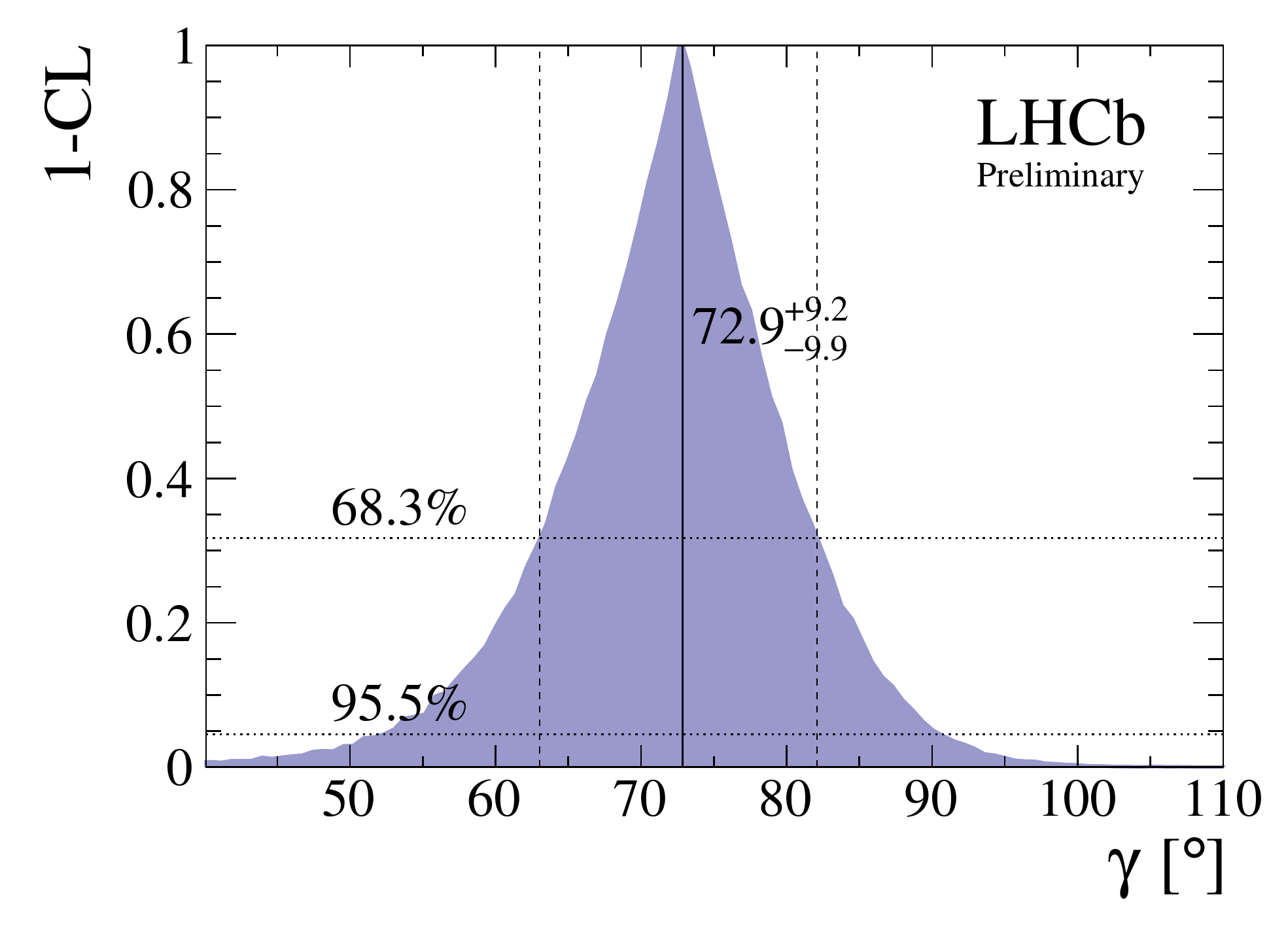}
\includegraphics[width=0.49\textwidth]{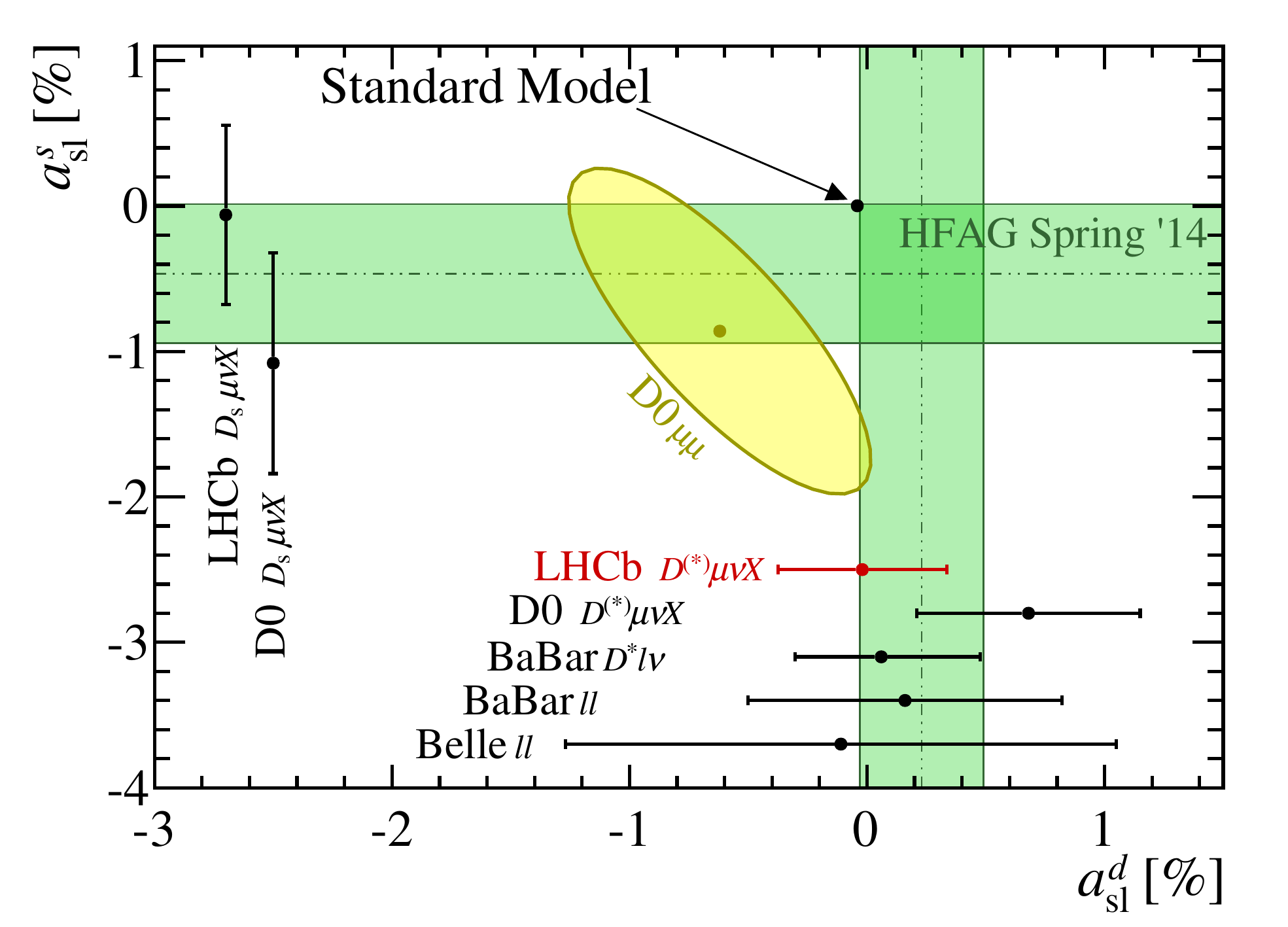}
\caption{(left) LHCb combination of \decay{\B}{D\kaon} decays measuring the 
  CKM phase $\gamma$~\cite{LHCb-CONF-2014-004}.
 (right) Experimental constraints on $A_\text{sl}^d$ and $A_\text{sl}^s$.}\label{Fig:2014-053}\label{Fig:CONF-2014-004}
\end{figure}
The $\gamma$ measurements of 
Refs.~\cite{LHCb-PAPER-2014-041,LHCb-PAPER-2014-038,LHCb-PAPER-2012-001,*LHCb-PAPER-2012-055,*LHCb-PAPER-2013-068,*LHCb-PAPER-2014-028} are 
then combined in an LHCb combination prepared for the CKM conference~\cite{LHCb-CONF-2014-004}.
Using only \decay{\B}{D\kaon} decay modes one finds $\gamma=(73\aerr{9}{10})\degrees$,
which is more precise than the corresponding combination of measurements from the \B factories~\cite{Bevan:2014iga}. The likelihood profile is shown in Fig.~\ref{Fig:CONF-2014-004} (left).

The same-sign dimuon asymmetry measured by the \dzero collaboration~\cite{Abazov:2013uma} 
and interpreted as a combination of the semileptonic asymmetries $A_\text{sl}^d$ and $A_\text{sl}^s$ in \Bd and \Bs decays, respectively, remains puzzling. The measured values
differ from the SM expectation by $3\sigma$. So far LHCb has not been able to confirm
or disprove this. The measurement from LHCb follows a different approach, 
looking at the \CP asymmetry
between partially reconstructed \decay{\B}{\Ds\mu\nu} decays, where the flavour of the 
\D identifies that of the \B. The measured value of 
$A_\text{sl}^s$~\cite{LHCb-PAPER-2013-033} and the
newly reported $A_\text{sl}^d$~\cite{LHCb-PAPER-2014-053} are both consistent 
with the SM and the \dzero value. The world average including measurements from the
\B factories and \dzero is not more conclusive. See Fig.~\ref{Fig:2014-053} (right).

Large \CP violation has also been found in charmless \bquark-hadron decays 
like \decay{\Bp}{h^+h^-h^+}~\cite{LHCb-PAPER-2014-044} ($h=\pion,\kaon$) and 
\decay{\Bp}{\proton\antiproton h^+}~\cite{LHCb-PAPER-2014-034}. In the former
case \CP asymmetries, integrated over the phase-space,
ranging from $-12\%$ (\decay{\Bpm}{\pipm\Kp\Km}) to 
$+6\%$ (\decay{\Bpm}{\pipm\Kp\Km}) are measured. Particularly striking 
features of these decays are very the large asymmetries in small
regions of the phase-space not related to any known resonance, 
which are of opposite sign for \decay{\Bpm}{h^\pm\Kp\Km}
and \decay{\Bpm}{h^\pm\pip\pim} decays. These could be a sign of long-distance
$\pip\pim\leftrightarrow\Kp\Km$ rescattering.

Finally, another important field is the study of \CP violation in beauty baryons.
The \Lb hadronisation fraction was measured to be surprisingly large at the 
LHC in the forward region~\cite{LHCb-PAPER-2014-004}, almost half of that
of \Bd mesons. These baryons can be used for measurements of \CP violation
with better precision than \Bs mesons. Searches have been performed by LHCb
with the decays \decay{\Lb}{\jpsi\proton\pim}~\cite{LHCb-PAPER-2014-020},
\decay{\Lb}{\Kz\proton\pim}~\cite{LHCb-PAPER-2013-061},
and by CDF with \decay{\Lb}{\proton h^-}~\cite{Aaltonen:2011qt}.
It is to be noted that to date no evidence of \CP violation in any
decay of a baryon has ever been reported.

\section{\decay{\Bs}{\mumu} and \decay{\Bd}{\mumu}}\label{Sec:Bmm}
The rare decay \decay{\Bs}{\mumu} proceeds in the SM by a box-type
annihilation diagram involving the \W boson and the \tquark quark. It is furthermore
helicity-suppressed. The most recent standard model prediction of its branching
fraction is $(3.66\pm0.23)\times10^{-9}$,~\cite{Bobeth:2013uxa},
where the uncertainty is dominated about equally 
by CKM matrix elements and the \Bs decay constant.
In this calculation the branching fraction is evaluated as an average 
over all decay times, see Refs.~\cite{DeBruyn:2012wj,*DeBruyn:2012wk} for more details.

The decay \decay{\Bs}{\mumu} has been searched for over three decades, with most recent
results from the 
Tevatron \cite{Aaltonen:2013as,*Abazov:2013wjb} and the 
LHC~\cite{LHCb-PAPER-2011-004,*LHCb-PAPER-2011-025,*Chatrchyan:2011kr,*LHCb-PAPER-2012-007,*Chatrchyan:2012rga,*ATLAS-CONF-2013-076}. The first evidence was reported in Summer 
2012 by LHCb~\cite{LHCb-PAPER-2012-043} using 2\:\invfb from the 2011 and half of the 2012 data.
A year later LHCb and CMS updated their results to the full Run I
 dataset~\cite{LHCb-PAPER-2013-046,*Chatrchyan:2013bka}. 

The data-sets in these two publications were then combined in a joint fit to 
the data of both experiments~\cite{LHCb-PAPER-2014-049}. 
\begin{figure}[!bt]\centering
\includegraphics[height=0.215\textheight]{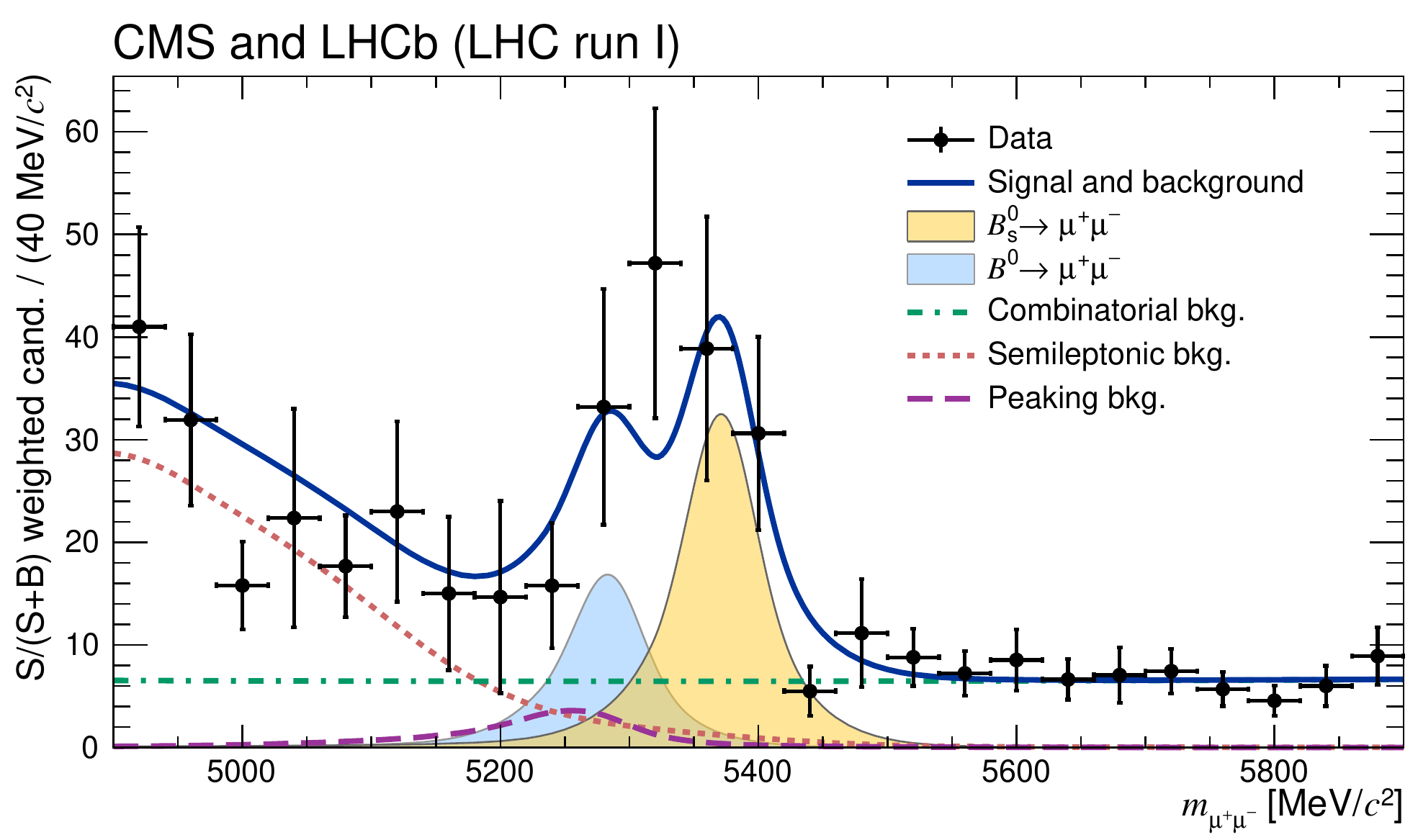}\hskip 0.02\textwidth
\includegraphics[height=0.215\textheight]{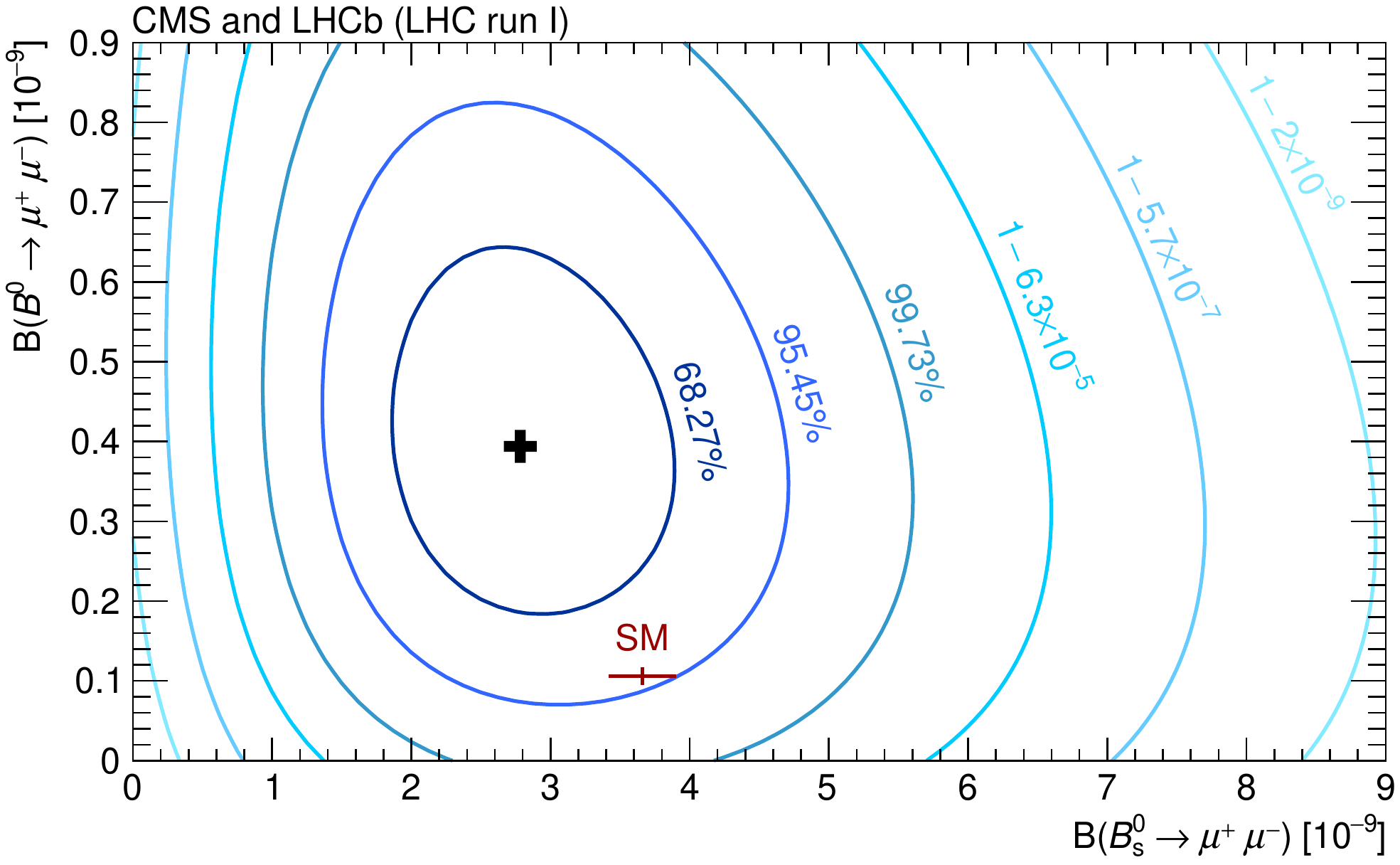}
\caption{(left) Example mass distribution from the combined LHCb and CMS dataset. (right)
Confidence level contours of the \Bsmm and \Bdmm measured branching fractions. The SM expectation is shown for comparison. Figures from Ref.~\cite{LHCb-PAPER-2014-049}.}\label{Fig:2014-049}
\end{figure}
This is the first time such a joint fit is done at the LHC.
The selections (and thus the data-sets) are left unchanged, but the fit models 
are aligned to be based on the same assumptions. Notably the treatment of the 
\decay{\Lb}{p\mup\nu} background was different between the two experiments,
and the effect of the lifetime acceptance on the admixture of heavy and light \Bs
states had been neglected in the CMS publications. As for the individual publications,
the fits are performed in each bin of boosted-decision-tree output. An example fit is shown 
in Fig.~\ref{Fig:2014-049} (left). The result of the 
combination is a clear first observation of the \Bsmm decay
and a $3\sigma$ excess over the background for the \Bdmm decay. 
In the latter case, the significance obtained from
Wilk's theorem is $3.2\sigma$, while a likelihood scan following the Feldman-Cousins 
method~\cite{prd573873} yields $3.0\sigma$. This is a deviation from the SM expectation
by $2.2\sigma$. 
The measured \Bdmm and \Bsmm branching fractions are compared to the SM expectation
in Fig.~\ref{Fig:2014-049} (right). More data from Run II and III will tell if the 
excess of \Bdmm is a statistical fluctuation or the indication of new physics.

\section{Rare electroweak decays}\label{Sec:blls}
The family of decays \decay{\bquark}{\squark\ellp\ellm} is a laboratory of new physics on its own. In particular the exclusive decay \decay{\Bz}{\Kstarz\ellp\ellm} ($\ell=e,\mu$) provides a very rich set of observables with different sensitivities to new physics and for which the theoretical predictions are available and affected by varying levels of hadronic uncertainties. In the case of some ratios of observables most of these uncertainties cancel, thus providing a clean test of the SM~\cite{Ali:1999mm,*Kruger:2005ep,*Altmannshofer:2008dz,*Egede:2008uy,*Bobeth:2008ij,Descotes-Genon:2013vna}. 

The differential decay width with respect to the dilepton mass squared $q^2$, 
the well-known forward-backward asymmetry $A_\text{FB}$, and the longitudinal polarisation fraction $F_\text{L}$ of the \Kstar resonance have been measured by many experiments~\cite{Aubert:2006vb,*Wei:2009zv,*Aaltonen:2011ja,*Chatrchyan:2013cda,*ATLAS-CONF-2013-038,LHCb-PAPER-2013-019} with no significant sign of deviations
from the SM expectation. 

\begin{figure}[tb]
\includegraphics[width=0.49\textwidth]{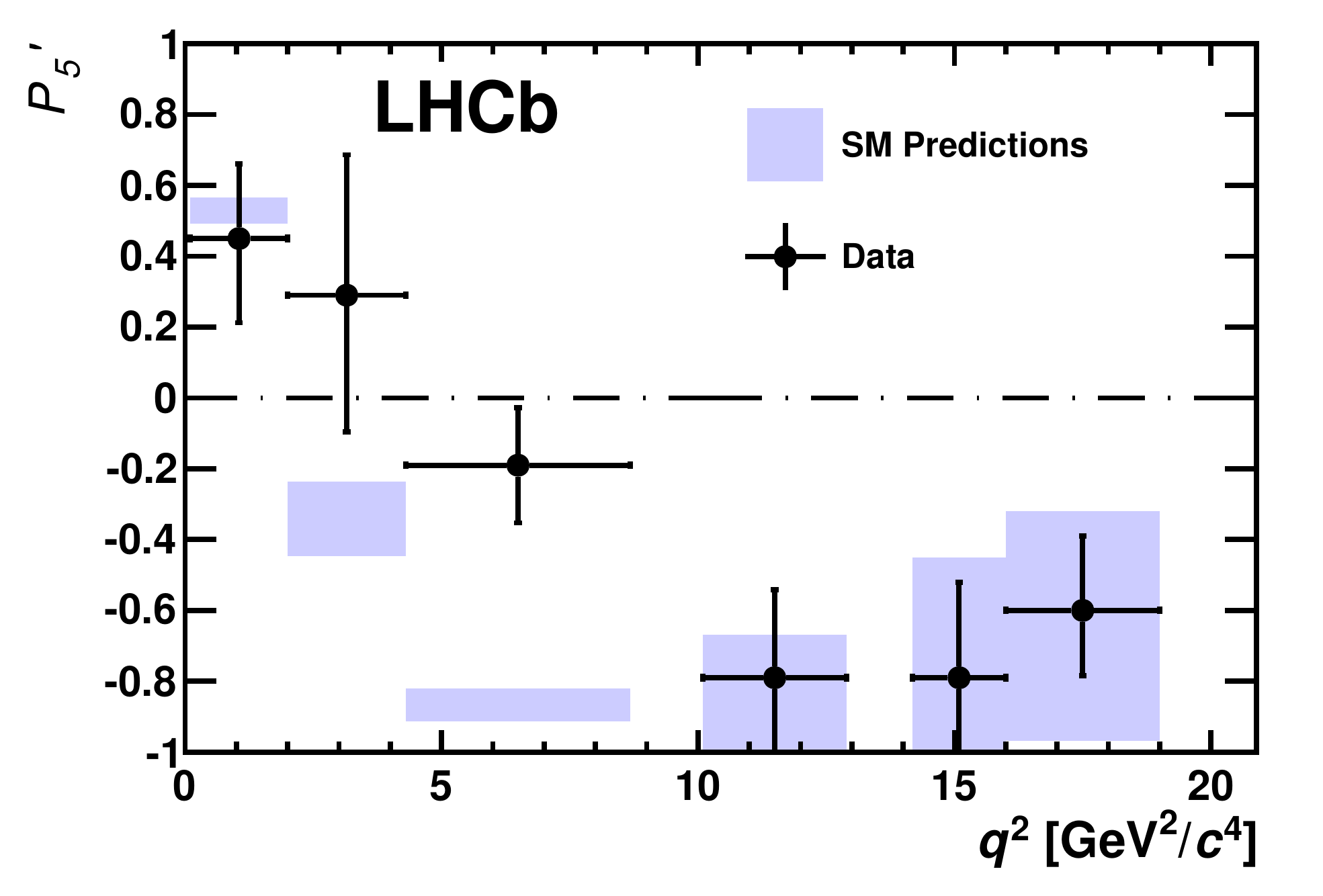}\hskip 0.01\textwidth
\includegraphics[width=0.49\textwidth]{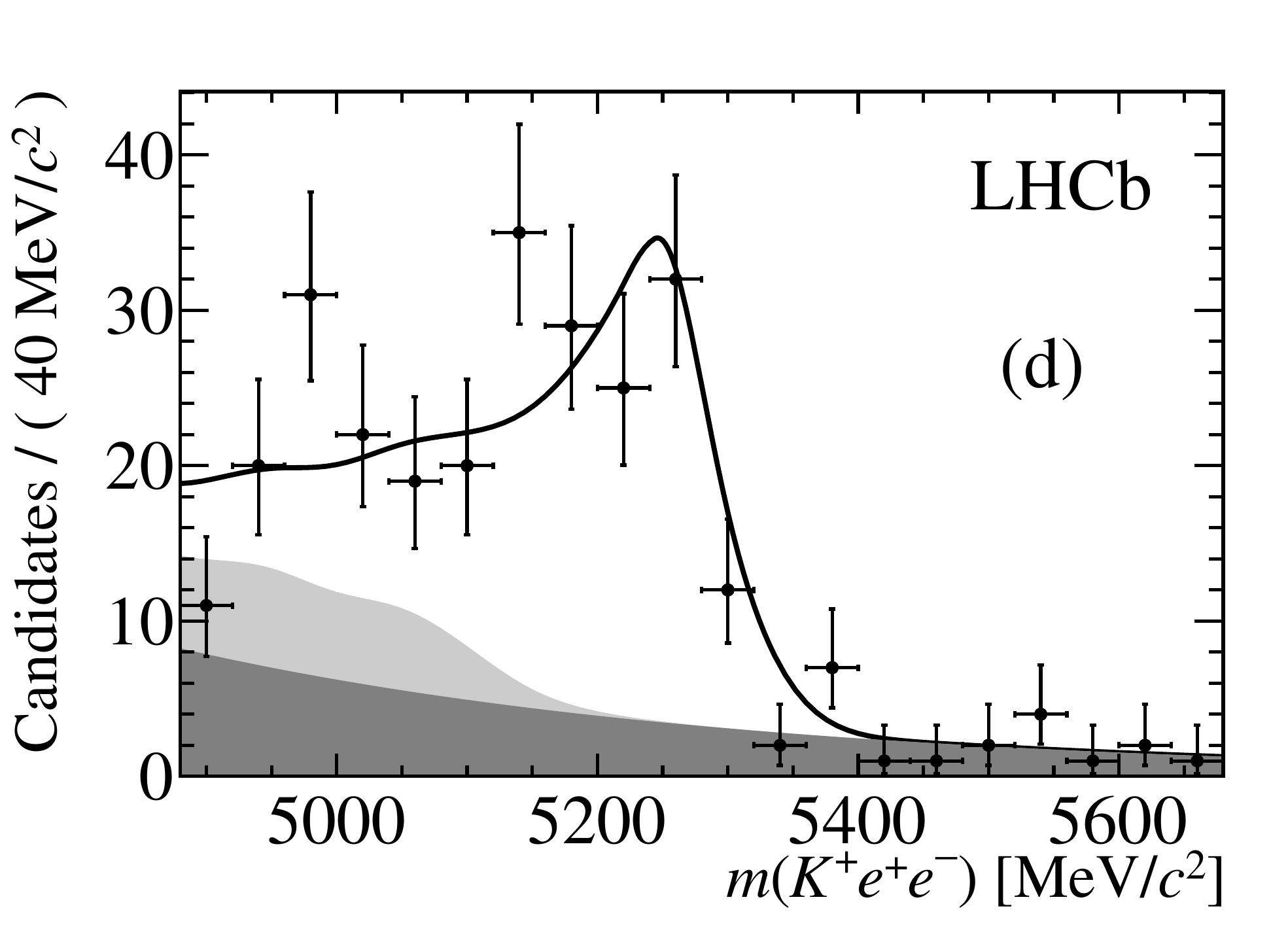}
\caption{(left) LHCb measurement of $P_5'$~\cite{LHCb-PAPER-2013-037},
(right) mass peak of \decay{\Bp}{\epem\Kp} for event triggered by one of the signal electrons 
at hardware level~\cite{LHCb-PAPER-2014-024}.}\label{Fig:2014-037}\label{Fig:2014-024}
\end{figure}

In a second analysis of the already published~\cite{LHCb-PAPER-2013-019} 2011 data, 
LHCb published another set of angular observables~\cite{LHCb-PAPER-2013-037}
suggested by Ref.~\cite{Descotes-Genon:2013vna}. In particular a $3.7\sigma$
local deviation of the $P_5'$ observable from the Standard Model expectation was observed in one bin of $q^2$, shown in Fig.~\ref{Fig:2014-037} (right). 

This measurement triggered a lot of interest in the theory community, with interpretation articles being very quickly submitted to journals. See Refs.~\cite{Gauld:2013qja,*Descotes-Genon:2013wba,*Altmannshofer:2013foa,*Datta:2013kja,*Mahmoudi:2014mja} for a small subset. It is not clear if this discrepancy is an experimental fluctuation, is due to under-estimated form factor uncertainties (See Ref.~\cite{Beaujean:2013soa}), or the manifestation of a heavy $Z'$ boson, among many other suggested explanations. The contribution
of \cquark{}\cquarkbar resonances is also being questioned~\cite{Lyon:2014hpa}
after the LHCb observation of \decay{\Bp}{\psi(4160)\Kp} with 
\decay{\psi(4160)}{\mumu}~\cite{LHCb-PAPER-2013-039}, 
where the $\psi(4160)$ and its interference with the non-resonant component 
accounts for 20\% of the rate for dimuon masses 
above 3770\:\mevcc. Such a large contribution was not expected.


Given the hint of abnormal angular distributions, LHCb tried to look for 
other deviations in several asymmetry measurements. The \CP asymmetry in 
\decay{\Bd}{\Kstar\mumu} and \decay{\Bpm}{\Kpm\mumu} turns out to be compatible with zero as expected~\cite{LHCb-PAPER-2014-032}, as does the isospin asymmetry between \decay{\Bd}{K^{(*)0}\mumu}
and \decay{\Bu}{K^{(*)+}\mumu}~\cite{LHCb-PAPER-2014-006}. The lepton universality
$R_K=\frac{{\cal B}(\decay{\Bp}{\Kp\mumu})}{{\cal B}(\decay{\Bp}{\Kp\epem})}$
is measured to be $0.745\aerr{0.090}{0.074}\pm0.036$~\cite{LHCb-PAPER-2014-024} in the $1<q^2<6\:\gevcc$ range, which indicates a $2.6\sigma$ tension with unity. 
Unlike at the \B factories, where electrons and muons contribute equally to \decay{b}{s\ellell} modes, at hadron colliders only muonic modes are normally used. 
Yet, LHCb has demonstrated its ability to use 
rare decays to electrons~\cite{LHCb-PAPER-2013-005}. The use of electrons is difficult due to the lower trigger efficiency and the poorer mass resolution due to bremsstrahlung. Figure~\ref{Fig:2014-024} shows the $\Kp\epem$ mass distribution for candidates with $1<q^2<6\:\gevcc$, which is 
safely far from the radiative tail of the \decay{\jpsi}{\epem} decay. 
This result is being interpreted as an indication of a new vector current ($Z'$?) 
that would couple
more strongly to muons and interfere destructively with the SM vector 
current~\cite{Hiller:2014yaa,*Ghosh:2014awa}.

\section{Conclusions}
The LHC is the new \bquark-hadron factory and will be dominating flavour physics until
the start of Belle II, and beyond in many decay modes. Atlas, CMS and LHCb
have presented interesting new results in rare decays, that set strong constraints 
on models beyond that SM and exhibit some discrepancies with the SM
predictions. It is not yet clear whether these are fluctuations, poorly understood 
form factors, or new physics. The upcoming Run II, with its higher centre-of-mass energy translating
into higher \bquark{}\bquarkbar cross-sections, may tells us.

\setboolean{inbibliography}{true} 
\bibliographystyle{LHCbNoTitle}    
\bibliography{LHCb-PAPER,LHCb-CONF,theory,exp}

\end{document}